\documentclass[prb,twocolumn,showpacs,superscriptaddress]{revtex4}
\usepackage{amsmath}
\usepackage{amssymb}
\usepackage{graphicx}
\usepackage{longtable}%
\usepackage{dcolumn}
\usepackage{bm}
\usepackage{color}
\begin{document}
\title
{First-order structural transition in the magnetically ordered phase of
Fe$_{1.13}$Te}
\author{S. R\"o{\ss}ler}
\email{roessler@cpfs.mpg.de}
\affiliation{Max Planck Institute for Chemical Physics of Solids,
N\"othnitzer Stra\ss e 40, 01187 Dresden, Germany}
\author{Dona Cherian}
\affiliation{Department of Physics, C.V. Raman Avenue, Indian
Institute of Science, Bangalore-560012, India}
\author{W. Lorenz}
\affiliation{Institut f\"ur Festk\"orperphysik, Technische
Universit\"at Dresden, 01062 Dresden, Germany}
\author{M. Doerr}
\affiliation{Institut f\"ur Festk\"orperphysik, Technische Universit\"at Dresden,
01062 Dresden, Germany}
\author{C. Koz}
\affiliation{Max Planck Institute for Chemical Physics of Solids,
N\"othnitzer Stra\ss e 40, 01187 Dresden, Germany}
\author{C.~Curfs}
\affiliation{ESRF, 6 Rue Jules Horowitz, BP 220, 38043 Grenoble Cedex 9, France}
\author{Yu. Prots}
\affiliation{Max Planck Institute for Chemical Physics of Solids,
N\"othnitzer Stra\ss e 40, 01187 Dresden, Germany}
\author{U. K. R\"o{\ss}ler}
\affiliation{IFW Dresden, Postfach 270016, 01171 Dresden, Germany}
\author{U. Schwarz}
\affiliation{Max Planck Institute for Chemical Physics of Solids,
N\"othnitzer Stra\ss e 40, 01187 Dresden, Germany}
\author{Suja Elizabeth}
\affiliation{Department of Physics, C.V. Raman Avenue, Indian
Institute of Science, Bangalore-560012, India}
\author{S. Wirth}
\affiliation{Max Planck Institute for Chemical Physics of Solids,
N\"othnitzer Stra\ss e 40, 01187 Dresden, Germany}
\date{\today}
%
%
%
\begin{abstract}
Specific heat, resistivity, magnetic susceptibility, linear
thermal expansion (LTE), and high-resolution synchrotron X-ray powder
diffraction investigations of single crystals Fe$_{1+y}$Te
(0.06 $\le y \le$ 0.15) reveal a splitting of a single,
first-order transition for $y \le$ 0.11 into two transitions
for $y \ge$ 0.13. Most strikingly, all measurements on identical
samples Fe$_{1.13}$Te consistently indicate that, upon cooling,
the magnetic transition at $T_{\rm N}$ precedes the first-order
structural transition at a lower temperature $T_{\rm s}$. The structural 
transition in turn coincides with a change in the character of the magnetic structure. The LTE measurements along the crystallographic $c$-axis displays a small distortion close to $T_{\rm N}$ due to a lattice striction as a consequence of magnetic ordering, and a much larger change at $T_{\rm s}$. The lattice symmetry changes, however, only below $T_{\rm s}$ as indicated by powder X-ray diffraction.
%
%
This behavior is in stark contrast to the sequence in which the phase
transitions occur in Fe pnictides.

\end{abstract}
\maketitle
\section{Introduction}
The recent discovery~\cite{kamihara2008iron} of superconductivity in iron-based
LaFeAsO$_{1-x}$F$_{x}$ (a member of the so-called 1111 family) 
with a transition temperature
$T_c$~=~26~K ignited tremendous
experimental and theoretical interest surrounding this family of
compounds. The superconducting members of the ferropnictides
exhibit transition temperatures as high as 56~K.
\cite{wang2008thorium} These materials display a strong
competition between structural, magnetic, and superconducting
transitions. One of the features common to both the high-$T_c$
copper oxide and Fe-based superconductors is that the
superconductivity emerges when an antiferromagnetic order is
suppressed by chemical substitution or doping. While the
initial interest was driven by the discovery of
superconductivity in different crystal systems with higher $T_c$,
the current emphasis is focused on understanding the origin of the
magnetic order and its relation to the superconductivity.
The members of the 1111 family undergo a
structural transition ($T_s$) followed by a magnetic transition ($T_{\rm N}$) at lower
temperatures, \cite{de2008magnetic,PhysRevLett.101.077005, luo2009evidence, Jesche2010} whereas in the 122 compounds these two
transitions occur simultaneously. \cite{PhysRevB.78.020503,
PhysRevB.78.180504, PhysRevB.78.014523} Several theoretical
models have been proposed for the possible microscopic mechanisms
coupling the magnetic and structural transitions. Among
these, the prominent ones are the itinerant-electron model
based on the nesting properties of the Fermi surface,
\cite{PhysRevB.79.134510} the local moment $J_{1}J_{2}$ model
favoring a spin-nematic-driven structural transition,
\cite{fang2008theory,xu2008ising} and an implementation of
orbital ordering into the double exchange model
\cite{PhysRevB.80.224504} where the structural transition is
induced by an orbital, rather than a magnetic, ordering. On the
other hand, a phenomenological Ginzburg-Landau model shows that
the magneto-elastic coupling between the different order parameters
can explain some of the experimentally observed phase transition
scenarios. \cite{cano2010interplay} However, none of the
theories developed up to now predict a possibility of a
structural transition taking place well within the
magnetically ordered phase, {\it i.e.} for magnetic ordering
occurring at a higher temperature than the structural transition,
$T_{\rm N} > T_s$, in Fe pnictides and chalcogenides.

Tetragonal Fe$_{1+y}$Te, the non-superconducting phase, 
occurs only with excess iron in the range
0.06 $\le y \le$ 0.15. Its crystal structure is intermediate between
the PbO ($B$10) and Cu$_{2}$Sb ($C$38) types. It may be
regarded as either PbO type with less than 0.2 extra atoms per cell, or
Cu$_{2}$Sb with more than 0.8 unoccupied Fe sites,
\cite{okamoto1990fe} both described within the $P4/nmm$ space
group. The nature of the antiferromagnetism in this material is
remarkably different in comparison to the FeAs superconductors. In
the FeAs based systems, the propagation vector of the
spin-density-wave (SDW) is ($\pi, \pi$) with respect to the
tetragonal lattice. This corresponds to a wave vector connecting
the $\Gamma$ and $M$ points in the Brillouin zone. In
Fe$_{1+y}$Te, in contrast, the corresponding wave vector is
($\pi$, 0), {\it i.e.} it is rotated by 45$^\circ$ with respect
to the ordering in the FeAs families. \cite{PhysRevLett.102.247001,
PhysRevB.79.054503, Liu2010} This implies that in Fe$_{1+y}$Te the nesting
properties of the Fermi surface do not play any role in the origin
of the antiferromagnetism. However, up to now it has been believed
that, in similarity to the 122 family, the antiferromagnetic
transition at around $T_{\rm N} = 68$~K is simultaneously
accompanied by a first-order structural distortion to the
monoclinic space group $P2_{1}/m$.~\cite{PhysRevB.79.054503} With
increasing amount of interstitial Fe the wave vector changes to an
incommensurate ($\delta \pi$,~0), and the crystal structure adopts
a higher symmetry (orthorhombic space group $Pmmn$) at low
temperatures. \cite{PhysRevLett.102.247001} Moreover, a very
recent neutron scattering experiment \cite{rodriguez2011magnetic}
revealed that the magnetic structure in Fe$_{1+y}$Te is even
more complex: at a critical concentration $y=$ 0.124, the magnetic
structure turns into an incommensurate helix.

Here, we report on concerted investigations on Fe$_{1+y}$Te
with 0.06 $\le y \le$ 0.15, with focus on samples with $y =~$ 0.11 and 0.13, {\it i.e.} around the
critical concentration. It is found that the $T_{\rm N}$ decreases
systematically from 70~K for $y =$ 0.06 to 57 K for $y =$ 0.11. Most importantly, we demonstrate that the magnetic and structural transitions in
Fe$_{1.13}$Te are separated in temperature by about 11~K, with the
antiferromagetic ordering occurring at higher temperature than the
structural transition. Our results show that the magnetic behavior
in the pnictides and chalcogenides could be entirely different and
cast a serious challenge to many existing theories of
Fe superconductors.
\begin{figure}[tb]
\centering \includegraphics[width=8.5 cm,clip]{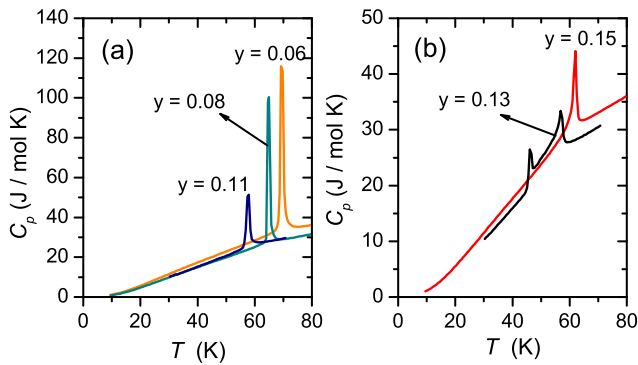}
\caption{Specific heat $C_{p}(T)$ of Fe$_{1+y}$Te for (a) y = 0.06, 0.08, 0.11, and (b) y = 0.13 and 0.15.}
\label{fig1}
\end{figure}
\begin{figure}[tb]
\centering \includegraphics[width=8.5 cm,clip]{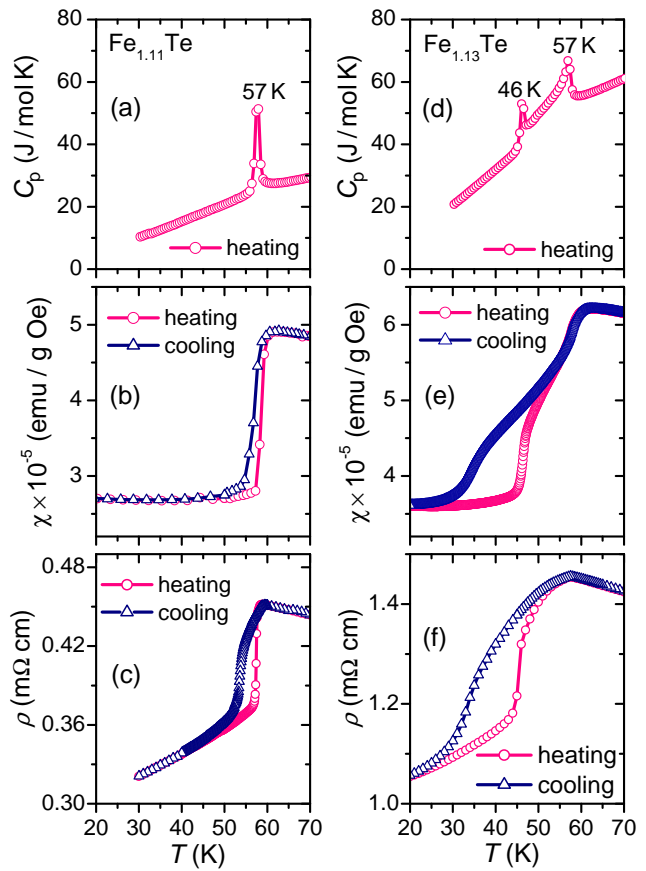}
\caption{Specific heat, dc magnetic susceptibility, and
resistivity (from top to bottom) of Fe$_{1.11}$Te (left) and
Fe$_{1.13}$Te (right) single crystals. The magnetic susceptibility
was measured in a field of 0.1 T applied along the $ab$-plane.}
\label{fig2}
\end{figure}

\section{Experimental details}
Single crystals Fe$_{1+y}$Te were grown using a horizontal
Bridgman setup. The details of the crystal growth procedure were
similar to those described in Ref.~[20], except for a different cooling rate. In the present case of Fe$_{1+y}$Te, after furnace translation, the samples were cooled down at a rate of 5 $^{\circ}$C/h from 950~$^{\circ}$C to 700~$^{\circ}$C, followed by cooling to room temperature with 25 $^{\circ}$C/h. The crystals were
characterized by Laue photographs, powder X-ray diffraction, chemical analysis, and
EDXS. The composition $y$ was
determined from the lattice parameters \cite{Koz} calibrated for mass loss
during growth, and EDXS. The specific heat $C_{p}(T)$ was measured
using a Quantum Design physical property measurement system (PPMS) with
a heat-pulse relaxation technique. At each measured temperature data point, 2~\% temperature rise and a measurement time with two time constants were used. 
The electrical resistivity $\rho(T)$ was also measured using the PPMS whereas the magnetic susceptibility $\chi(T)$ was obtained by means of a SQUID
vibration sample magnetometer. The diffraction data were
collected on the high-resolution powder diffraction beamline ID31
(wavelength of 39.992~pm) at the ESRF, Grenoble, using a
liquid-He cryostat. A sensitive tilted-plate
capacitive dilatometer \cite{rotter1998miniature} with a
resolution of relative length changes $\Delta l/l_{0} = 10^{-7}$ was
employed to measure the linear thermal expansion (LTE) along the crystallographic
$c$ axis.

\section{Results and discussion}
The temperature dependence of specific heat $C_p(T)$ of Fe$_{1+y}$Te for $y$~=~0.06, 0.08, 0.11, 0.13, and 0.15 is presented in Fig.~\ref{fig1}. For $y$~=~0.06, $C_p$ shows a sharp first-order peak corresponding to a simultaneous magnetic and structural transition at $T_{\rm N}=T_{\rm s}\sim$~70~K. This transition temperature 
monotonically decreases to 57~K with $y$ increasing to 0.11. The transition temperature is drastically suppressed with respect to
that in Fe$_{1.06}$Te due to the increased amount of interstitial Fe,
a finding consistent with previous studies.
\cite{PhysRevB.80.174509} Instead of a continued suppression of $T_{\rm N}$, the further increased amount of interstitial Fe in Fe$_{1.13}$Te gives rise to a dramatically different behavior, Fig.~\ref{fig1}(b). Two clearly distinct transitions are observed: First, a $\lambda$-like transition at 57~K, followed by a first-order phase transition at 46 K. For even more interstitial Fe, $y$~=~0.15, the temperature at which the $\lambda$-like transition occurs increases to 63~K, but the low-temperature first-order transition could not be clearly resolved in the specific heat measurements.

To investigate the nature of phase transitions in Fe$_{1+y}$Te around the critical concentration where successive phase transitions occur, we conducted magnetic susceptibility $\chi(T)$ and electrical resistivity $\rho(T)$ measurements for $y$ = 0.11 and 0.13. In Fig.~\ref{fig2}, $\chi(T)$ and $\rho(T)$ data are presented along with $C_p(T)$ for these samples for comparison. $C_p(T)$ in Fig.~\ref{fig2}(a) clearly displays a sharp first-order transition at 57~K. This temperature corresponds to a simultaneous structural and SDW transition, $T_s \approx T_{\rm N}$. The coincident SDW
transition is confirmed by a sudden decline in the magnetic
susceptibility $\chi(T)$ around 57~K, Fig.~\ref{fig2}(b). The
cooling and heating susceptibility cycles measured in a field of
0.1~T show a thermal hysteresis typical for a first-order phase
transition. The in-plane ($ab$ plane) resistivity $\rho(T)$ displays a
corresponding transition from a semiconducting to metallic
behavior, as can be seen in Fig.~\ref{fig2}(c). 
\begin{figure}[tb]
\centering \includegraphics[width=8.5 cm,clip]{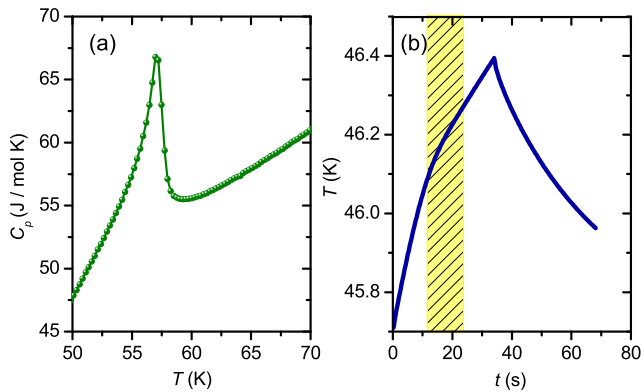}
\caption{(a) High temperature-resolution specific heat data for Fe$_{1.13}$Te showing a clear $\lambda$-like transition at 57~K. (b) The temperature-time relaxation curve depicting the arrests (shaded region) due to latent heat at the first-order phase transition at $T\sim$ 46~K in Fe$_{1.13}$Te.}
\label{fig3}
\end{figure}
In Fig.~\ref{fig2}(d), $C_p(T)$ for Fe$_{1.13}$Te is
given. 
The high-temperature transition at $T_{\rm N}$ = 57 K in
Fig.~\ref{fig2}(d) is unmistakable $\lambda$-like, proving that
the phase transition is of second order, and can be described by a
power law divergence. The $\lambda$-like  shape of this transition is obvious in the high temperature-resolution data provided in Fig. \ref{fig3}(a).
The transition at lower temperature, $T_s =
46$~K, in Fe$_{1.13}$Te  is similar in shape to the single one observed in
Fe$_{1.11}$Te (see Fig.~\ref{fig2}(a)), where a latent heat is
involved and hence, is of first-order in nature. In order to confirm the first-order nature of the transition, the temperature-time relaxation curve is presented in Fig. \ref{fig3}(b). It clearly displays a temperature arrest around 46~K in the warming part (shaded region) of the relaxation curve. As expected for this case, it clearly displays a first-order nature of the transition. In the cooling part, however, the temperature arrest was not observed, suggesting that the first-order transition is spread over a wider temperature interval. This argument is also supported by a broader transition observed in the $\chi(T)$ cooling measurement, Fig.~\ref{fig2}(e). This finding, along with the results presented below, indicates
that a change in the crystal structure occurs only at the
low-temperature transition. With this assignment the lower
values of $C_p$ just below $T_s$ suggest a discontinuous jump in
the phononic background. This typically happens at a structural
transition due to softened optical phonons.

A continuous transition followed by a first-order transition in
Fe$_{1.13}$Te can also be discerned from the susceptibility
measurements shown in Fig.~\ref{fig2}(e): $\chi(T)$ first
decreases continuously at around 57 K, followed by a sharp jump
upon further reduction of temperature. A huge thermal
hysteresis covering a width of $\approx$~25~K in the field-cooled and
heating protocols was found only at the low temperate transition.
However, the thermal hysteresis in the resistivity is broader and
remains up to the high temperature transition,
Fig.~\ref{fig2}(f). 
Here we note that
similar successive phase transitions have also been observed
in resistivity measurements on Fe$_{1.086}$Te above an
applied pressure of about 1~GPa. \cite{okada2009successive} This
suggests that the addition of interstitial Fe beyond $y
> 0.11$ produces similar effects like application of pressure
on Fe$_{1+y}$Te with smaller amount $y$ of interstitial Fe.

\begin{figure}[tb]
\centering \includegraphics[width=8 cm,clip]{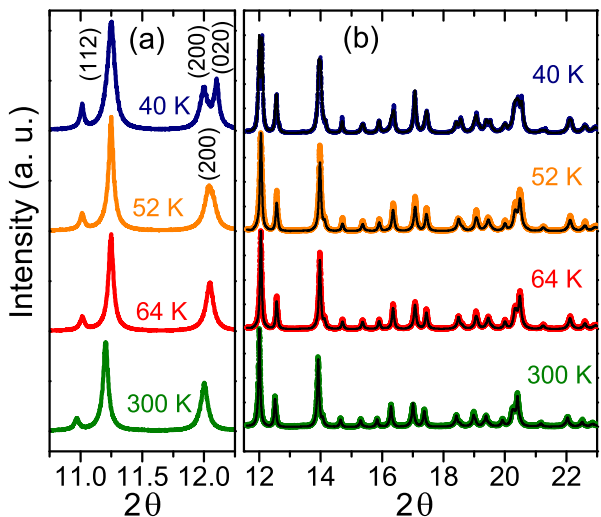}
\caption{Synchrotron X-ray diffraction of the powdered
Fe$_{1.13}$Te single crystal at temperatures above $T_{\rm N} =
57$~K, between $T_{\rm N}$ and $T_s$, and below $T_s = 46$~K,
respectively. (a) Zoomed range $10.75^{\circ} \le 2\theta \le
12.25^{\circ}$. The single (112) peak  and the splitting of the
(200) tetragonal peak into two peaks (200) and (020) at $T = 40~K
< T_{s}$ indicate an orthorhombic low-temperature structure. (b)
Overview spectra $11.5^{\circ} \le 2\theta \le 23^{\circ}$. The
black lines represent fitted curves.} \label{fig4}
\end{figure}

In order to unambiguously confirm the above assignment of the
structural transition to $T_s = 46$~K in Fe$_{1.13}$Te we
performed high-resolution synchrotron X-ray diffraction of the
powdered single crystals at several selected temperatures. In
Fig.~\ref{fig4}, the diffraction data collected above $T_{\rm N}$
(300~K, 64~K), between $T_{\rm N}$ and $T_s$ (52~K), and below
$T_s$ (40~K) are presented. The pattern at 300~K could be well
fitted within the tetragonal $P4/nmm$ space group,
\cite{PhysRevLett.102.247001, PhysRevB.79.054503} see
Fig.~\ref{fig4}(b). Upon crossing $T_{N}$ = 57~K, the structure remains
unchanged, {\it i.e.} tetragonal, as can be seen from the pattern
at $T = 52$~K. Note that at 52~K, the peak corresponding to the
(200) reflection starts to broaden indicating a structural
instability arising due to the fluctuations in the vicinity of the
phase transition. Notably, the pattern at 40~K is clearly
different. The observed splitting of the diffraction peaks,
specifically of the tetragonal (200) at higher $T$ into (200) and
(020) at 40~K $[$see ~Fig.~\ref{fig4}(a)$]$, is compatible with an
orthorhombic symmetry below $T_s$. 
Moreover, the intensities observed at 40~K can be described
by a structure model in the space group $Pmmn$.
This result is consistent with neutron scattering studies
on Fe$_{1.141}$Te. \cite{PhysRevLett.102.247001}

\begin{figure}[t]
\centering \includegraphics[width=6.0cm,clip]{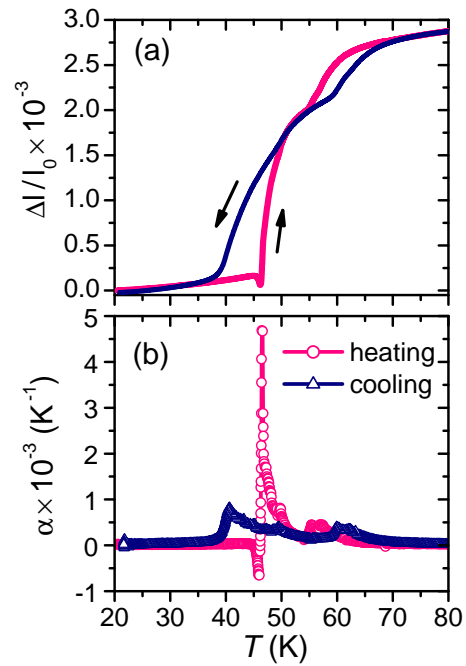}
\caption{(a) The linear thermal expansion (LTE) of the
Fe$_{1.13}$Te single crystal measured along the crystallographic
$c$ axis in both cooling and heating cycles. (b) The corresponding
coefficients of the LTE, $\alpha (T) = ({1}/{l_0})({dl}/{dT})$.}
\label{fig5}
\end{figure}

In Fig.~\ref{fig5}(a) and (b), we present the LTE $\varDelta l/l_0$ and the 
corresponding LTE coefficient $\alpha (T) = ({1}/{l_0})({dl}/{dT})$ of the
Fe$_{1.13}$Te sample along the crystallographic $c$ axis. Upon
cooling, the LTE first displays a shoulder at $T \sim 62$~K due to
the incommensurate antiferromagnetic order, followed by a
broadened jump at the structural transition. A discontinuity
corresponding to the latter transition is nicely resolved in
$\alpha (T)$ at $T \approx 41$~K, Fig.~\ref{fig5}(b). The
LTE curve very much resembles the one for $\chi(T)$ upon cooling
in Fig.~\ref{fig2}(e). For increasing temperature, $\varDelta
l/l_0$ displays a large thermal hysteresis and a sharp increase at
$T \approx 46$~K due to the structural transition. However,
the transition corresponding to the magnetic ordering in the
warming cycle occurs at around 57~K, {\it i.e.} at a significantly
lower temperature than the corresponding transition in the cooling
cycle. This kind of reverse hysteresis is rather unusual for a
continuous phase transition (see discussion below).

\begin{figure}[t!]
\centering \includegraphics[width=7 cm,clip]{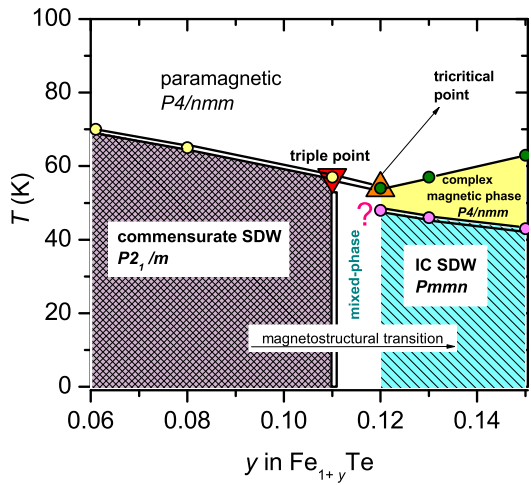}
\caption{A tentative schematic temperature-composition phase
diagram of Fe$_{1+y}$Te based on our specific heat, LTE, and X-ray diffraction results (circles, color online) as
well as on results of Refs. \cite{PhysRevLett.102.247001,
rodriguez2011magnetic}. The double lines depict a first-order
phase transition, the single line represents a continuous phase
transition. The origin of the low-temperature first-order line is
unclear, therefore marked by ``?''. IC SDW denotes an
incommensurate spin-density wave.} 
\label{fig6}
\end{figure}

Combining our results with those from neutron scattering
experiments \cite{PhysRevLett.102.247001,rodriguez2011magnetic}
allows a detailed interpretation of the magnetic and structural
transitions in Fe$_{1+y}$Te. For low Fe concentrations
within the range $0.06<y<0.11$, the structural
transition assigned to the monoclinic $P2_1/m$ space group appears to take
place simultaneously with magnetic ordering as a first-order
magneto-elastic transition, as depicted in the schematic phase
diagram Fig.~\ref{fig6}.
%
%
%
%
%
For Fe$_{1.124}$Te, an incommensurate helix is found with a
propagation vector $(q,0,\frac{1}{2})$ r.\ l.\ u.\ where $q
\sim 0.445$, which decreases to 0.4 with increasing temperature in
an intermediate range 40~K $< T< 57$~K.~\cite{rodriguez2011magnetic} Within the same temperature range, this helicoid and an incommensurate SDW with
different period $q \sim 0.38$ are observed simultaneously.
\cite{rodriguez2011magnetic} This is assigned as a complex
magnetic phase in Fig.~\ref{fig6}.
%
Symmetry considerations strongly restrict the possible phase
transition mechanism for the formation of the helimagnetic state.
The paramagnetic space group $P4/nmm$ comprises only
one-dimensional irreducible representations in the little
group for propagation vectors $(q,0,\frac{1}{2})$.
\cite{RevModPhys.40.359,tolédano1987landau} Thus, a continuous
transition from the paramagnetic into the helical state cannot
take place as a mode instability according to standard
Landau theory. \cite{DeGennes75} However, there are Lifshitz-type
invariants that couple different irreducible representations
and can (i) produce a helimagnetic state and (ii) give rise to the
nucleation of kink-like modulations of the basic helical
modulation on a mesoscopic length scale. \cite{Dz64,
DeGennes75, tolédano1987landau, Mukamel85} The existence of
Lifshitz-invariants in helimagnets with strong phase-amplitude
interaction causes unconventional magnetic ordering transitions
\cite{UKR11}. Anisotropic magnetic couplings can cause
continuous or discontinuous transitions. \cite{Mukamel85}
The presence of such anisotropies in Fe$_{1+y}$Te is suggested by
the lock-in transition at $y<0.11$ and the marked
magnetostructural transition in this composition range. Turner
$et~al.$ have proposed a model with localized Fe moment coupled to
an orbital order, \cite{PhysRevB.80.224504} which could provide a
microscopic mechanism for the magneto-elastic coupling that
should vary with $y$ in Fe$_{1+y}$Te. Thus, for $y>0.11$ a similar
locking-in of the helix may occur owing to possible anisotropic couplings.
But, this lock-in transition produces an incommensurate magnetic
ordering. This can be understood phenomenologically, if the free
energy of the structural order parameter contains dispersive
couplings. \cite{tolédano1987landau} Owing to the intrinsic
disorder by the partially occupied Fe sublattice, strong pinning
of the kink-like solitons, {\it i.e.} domain walls between regions
with SDW and helical order, will occur. The very wide hysteresis
in our LTE data is consistent with such a pinning. Thus, we
explain our results in Fe$_{1.13}$Te by a magnetostructural
transformation of the helix in the tetragonal lattice state into a
SDW with orthorhombic distortions subject to strong pinning. This
is also in line with the observation of co-existing SDW
and helicoidal order associated with different lattice structures
in Fe$_{1.124}$Te, which was detected by diffraction in
neutron scattering experiments.~
\cite{rodriguez2011magnetic} The transformation process can be
pictured as a decomposition of the magnetic helix into an
incoherent sequence of SDW sections with interspersed helicoidal
kinks on a {\em mesoscopic length scale}.

Features of two successive phase transition have been earlier observed
by Hu $et~al.$~\cite{Hu2009} in Fe$_{1.09}$Te and Fe$_{1.14}$Te. However,
they did not associate these transitions to individual magnetic and structural
phase transitions. Instead, their results were explained based on the nesting
properties of the Fermi surface: the higher level of excess Fe corresponds 
to larger size mismatch between cylindrical electron and hole Fermi surfaces.
The two transitions may then represent successive SDW Fermi-surface nestings
of separate electron-hole cylindrical pieces. However, neutron scattering studies~\cite{PhysRevLett.102.247001} indicated that the Fermi surface nesting does not play a role in the magnetic ordering in Fe$_{1+y}$Te. More recently, Zaliznyak $et~al.$~\cite{Zaliznyak} report $\chi(T)$
data for Fe$_{1.09}$Te which is very similar to those of our sample Fe$_{1.13}$Te. From the neutron scattering studies on Fe$_{1.1}$Te, Zaliznyak $et~al.$ first find a structural distortion at $T_{\rm s}$~=~63 K, followed by a magnetic ordering at $T_{\rm N}$ = 57.5(5)~K. At $T_{\rm m}\le $ 45 K, a lock-in transition was observed at an incommensurate wave-vector ($0.48, 0, \frac{1}{2}$). The (201) lattice Bragg peak in their studies does not show a clear splitting down to 9~K. From their data,\cite{Zaliznyak} it was not possible to distinguish between the monoclinic $P2_{1}/m$ and the orthorhombic $Pnmm$ structures. Since no phase transition features were observed either in the susceptibility or in the specific heat at 63~K, the structural distortion at $T_{\rm s}$~=~63 K was identified from the onset of broadening of (201) reflection. But, such a broadening may also be associated with a symmetry-preserving lattice striction caused by a strong magneto-elastic coupling close to $T_{\rm N}$.  This distortion is followed by a change in the crystal symmetry at the first-order lock-in transition at 45~K, as observed in our high resolution diffraction data. Thus, we propose that the structural transition occurs simultaneously at a temperature where a change in the character of magnetic structure also takes place well within the magnetically ordered phase.  

The symmetry changing structural phase-transition in the magnetically ordered phase is unusual in the parent compounds of Fe-superconductors. In the case of 1111 pnictides \cite{de2008magnetic,PhysRevLett.101.077005, luo2009evidence, Jesche2010} or partially Co substituted BaFe$_{2}$As$_{2}$ \cite{Chu2010, Pratt2009} and CaFe$_{2}$As$_{2}$,\cite{Chuang2010} where these two
phase transitions occur separately, the structural transition always takes place at a higher temperature than $T_{\rm N}$. Within an effective Heisenberg-type local-moment ($J_{1}$-$J_{2}$-$J_{z}$) model, this splitting arises as a consequence of Ising-like magnetic couplings \cite{xu2008ising, Fang2008, Qi2009} with a very weak interlayer interaction. In support of this theory, the temperature gap between  $T_{\rm s}$ and $T_{\rm N}$ was experimentally found to increase with increasing distance between FeAs layers.\cite{luo2009evidence} In the case of Fe$_{1+y}$Te, however, the $c$-lattice constant decreases with increasing $y$. \cite{PhysRevLett.102.247001, Koz} Yet the successive phase transitions occur for $y \ge 0.12$; 
\textit{i.e.}, for the compositions with shorter $c$-lattice constant. Thus, the microscopic mechanisms driving the  phase transitions in Fe$_{1+y}$Te seems to be fundamentally different from those in the case of Fe-arsenides.

\section{conclusions}
In conclusion, our thermodynamic, structural, and thermal
expansion data on Fe$_{1.13}$Te give clear evidence for
magnetic ordering taking place at {\em higher} temperatures
than the structural phase transition. The structural 
transition in turn coincides with a change in the nature of the magnetic structure.
This is exactly opposite to the behavior observed in the 1111-systems,
for which the magnetic transition occurs at lower
temperatures than the structural one. The exact nature
of the microscopic coupling mechanisms in Fe$_{1+y}$Te needs
to be explored further as it appears to be key in
understanding the interplay between localized and itinerant
magnetism as well as superconductivity in Fe chalcogenides.
As the pairing mechanism in both cuprates and Fe-superconductors probably
involves spin fluctuations, understanding the nature of magnetic interactions is of utmost importance.\\

\section{Acknowledgements}
We thank  C.~Geibel, Yu. Grin, Q.
Si, F.~Steglich, O.~Stockert, P.~Thalmeier, and L.~H.~Tjeng for
stimulating discussions. This work is partially supported by
DAAD - DST exchange program ID 50726385.


\end{document}